\documentclass[final]{svjour3}
\usepackage{gensymb}
\usepackage{graphicx}
\usepackage{rotating}
\usepackage{amssymb}
\usepackage{mathptmx}
\usepackage{comment}
\usepackage{floatrow}
\usepackage[numbers]{natbib}
\makeatletter
\journalname{Journal of Low Temperature Physics}

\bibpunct{}{}{,}{s}{}{,}

\begin{document}

\newcommand{\hdblarrow}{H\makebox[0.9ex][l]{$\downdownarrows$}-}
\title{Characterizing the Sensitivity of 40 GHz TES Bolometers for BICEP Array}

\author{C.~Zhang$^1$ \and P.~A.~R.~Ade$^2$ \and Z.~Ahmed$^{3,4}$ \and M.~Amiri$^5$ \and D.~Barkats$^6$ \and R.~Basu Thakur$^1$ \and C.~A.~Bischoff$^7$ \and J.~J.~Bock$^{1,8}$ \and H.~Boenish$^6$ \and E.~Bullock$^9$ \and V.~Buza$^6$ \and J.~Cheshire$^9$ \and J.~Connors$^6$ \and J.~Cornelison$^6$ \and M.~Crumrine$^9$ \and A.~Cukierman$^4$ \and M.~Dierickx$^6$ \and L.~Duband$^{10}$ \and S.~Fatigoni$^5$ \and J.~P.~Filippini$^{11,12}$ \and G.~Hall$^9$ \and M.~Halpern$^5$ \and S.~Harrison$^6$ \and S.~Henderson$^{3,4}$ \and S.~R.~Hildebrandt$^8$ \and G.~C.~Hilton$^{13}$ \and H.~Hui$^1$ \and K.~D.~Irwin$^{3,4}$ \and J.~Kang$^4$ \and K.~S.~Karkare$^{6,14}$ \and E.~Karpel$^4$ \and S.~Kefeli$^1$ \and J.~M.~Kovac$^6$ \and C.~L.~Kuo$^{3,4}$ \and K.~Lau$^9$ \and K.~G.~Megerian$^8$ \and L.~Moncelsi$^1$ \and T.~Namikawa$^{15}$ \and H.~T.~Nguyen$^8$ \and R.~O'Brient$^{8,1}$ \and S.~Palladino$^7$ \and N.~Precup$^9$ \and
T.~Prouv\'{e}$^{10}$\and C.~Pryke$^9$ \and B.~Racine$^6$ \and C.~D.~Reintsema$^{13}$ \and S.~Richter$^6$ \and A.~Schillaci$^1$ \and B.~Schmitt$^6$ \and R.~Schwarz$^6$ \and C.~D.~Sheehy$^{16}$ \and A.~Soliman$^1$ \and T.~St.~Germaine$^6$ \and B.~Steinbach$^1$ \and R.~V.~Sudiwala$^2$ \and K.~L.~Thompson$^{3,4}$ \and C.~Tucker$^2$ \and A.~D.~Turner$^8$ \and C.~Umilt\`{a}$^7$ \and A.~G.~Vieregg$^{14}$ \and A.~Wandui$^1$ \and A.~C.~Weber$^8$ \and D.~V.~Wiebe$^5$ \and J.~Willmert$^9$ \and W.~L.~K.~Wu$^{14}$ \and E.~Yang$^4$ \and K.~W.~Yoon$^4$ \and E.~Young$^4$ \and C.~Yu$^4$  }

\institute{$^1$Department of Physics, California Institute of Technology, Pasadena, California 91125, USA
\\$^2$School of Physics and Astronomy, Cardiff University, Cardiff, CF24 3AA, United Kingdom
\\$^3$Kavli Institute for Particle Astrophysics and Cosmology, SLAC National Accelerator Laboratory, 2575 Sand Hill Rd, Menlo Park, California 94025, USA
\\$^4$Department of Physics, Stanford University, Stanford, California 94305, USA
\\$^5$Department of Physics and Astronomy, University of British Columbia, Vancouver, British Columbia, V6T 1Z1, Canada
\\$^6$Harvard-Smithsonian Center for Astrophysics, Cambridge, Massachusetts 02138, USA
\\$^7$Department of Physics, University of Cincinnati, Cincinnati, Ohio 45221, USA
\\$^8$Jet Propulsion Laboratory, Pasadena, California 91109, USA
\\$^9$Minnesota Institute for Astrophysics, University of Minnesota, Minneapolis, 55455, USA
\\$^{10}$Service des Basses Temp\'{e}ratures, Commissariat \`{a} lEnergie Atomique, 38054 Grenoble, France
\\$^{11}$Department of Physics, University of Illinois at Urbana-Champaign, Urbana, Illinois 61801, USA
\\$^{12}$Department of Astronomy, University of Illinois at Urbana-Champaign, Urbana, Illinois 61801, USA
\\$^{13}$National Institute of Standards and Technology, Boulder, Colorado 80305, USA
\\$^{14}$Kavli Institute for Cosmological Physics, University of Chicago, Chicago, IL 60637, USA
\\$^{15}$Department of Applied Mathematics and Theoretical Physics, University of Cambridge, Wilberforce Road, Cambridge CB3 0WA, UK
\\$^{16}$Physics Department, Brookhaven National Laboratory, Upton, NY 11973
\\ \email{czzhang@caltech.edu}}

\maketitle

\begin{abstract}
The BICEP/Keck (BK) experiment aims to detect the imprint of primordial 
gravitational waves in the Cosmic Microwave Background polarization, 
which would be direct evidence of the inflation theory. While the 
tensor-to-scalar ratio has been constrained to be $r_{0.05}<0.06$ at 95\% c.l., 
further improvements on this upper limit are hindered by polarized 
Galactic foreground emissions and removal of gravitational lensing polarization. The 30/40 GHz receiver of the BICEP Array 
(BA) will deploy at the 
end of 2019 and will constrain the synchrotron foreground with 
unprecedented accuracy within the BK sky patch. We will show the design of the 
30/40\,GHz detectors and test results summarizing its performance. The low optical and 
atmospheric loading at these frequencies requires our TES detectors to 
have low saturation power in order to be photon-noise dominated. To 
realize the low thermal conductivity required from a 250 mK base temperature, we developed new bolometer leg designs. We will present the relevant measured detector parameters: $G$, $T_c$, $R_n$, $P_{sat}$, and spectral bands, and noise spectra. We achieved a per bolometer NEP including all noise components of  $2.07\times10^{-17}~{\rm W/\sqrt{Hz}}$, including an anticipated photon noise level $1.54\times10^{-17}~{\rm W/\sqrt{Hz}}$.

\keywords{Cosmology, B-Mode polarization, Inflation, CMB, TES}

\end{abstract}

\section{Introduction}

Inflation theory builds upon the Standard Cosmological Model to provide an explanation of the origin of structure in our universe. Many families of inflationary models predict the presence of primordial gravitational waves, which would leave imprints in the Cosmic Microwave Background (CMB) polarization during recombination. BICEP/Keck (BK) experiments are aiming to measure the degree-scale B-mode polarization pattern in the CMB, which could constrain the tensor-to-scalar ratio $r$ and place limits on the energy scale and potential of Inflation. 

The gravitational wave induced B-mode patterns are very faint- modern experiments constrain the tensor-scalar ratio to be $r_{0.05}<0.06$ at 95\% confidence level with $\sigma(r)=0.02$ [1]. These experiments must also constrain and remove B-modes from lensing of E-modes by large scale structure and the polarized emission from galactic foregrounds such as dust and synchrotron. Galactic foregrounds have a different spectral signature than the CMB, which allows component separation with multi-frequency observations. Keck receivers operating at 220 and 270\,GHz, added in 2015 and 2017 respectively, have provided the ability to constrain the dust foreground in our observing sky patch.

BICEP Array (BA) [2][3] is the new generation of BK experiments, which will observe through the atmospheric windows between 30 and 270\,GHz with a total of four receivers. The first receiver  will observe at 30 and 40\,GHz to constrain the synchrotron foreground. The receiver is designed to resolve $35<l<300$ features in a 600 deg$^2$ sky patch. The focal plane of BA will include twelve 6-inch detector modules.  Half of the first camera's focal plane modules will have 30\,GHz detectors while the others will have 40\,GHz detectors, providing total detector counts of 192 and 300 respectively. The detectors we are using in this first receiver at 30/40\,GHz are Transition Edge Sensor (TES) bolometers coupled with polarization-sensitive coplanar slot antennas. In this paper we will mainly focus on the detector development of 30/40\,GHz. The central engineering challenge for these low frequency detectors is realizing background-limited noise performance from a 250 mK base temperature, which requires a high degree of thermal isolation. See [5] for an overall update of first BA receiver and [6] for the detailed antenna design and beam characterization.

\section{Antenna and Bolometer Design}

\begin{figure}[htbp]
\begin{center}
\includegraphics[width=0.7\linewidth]{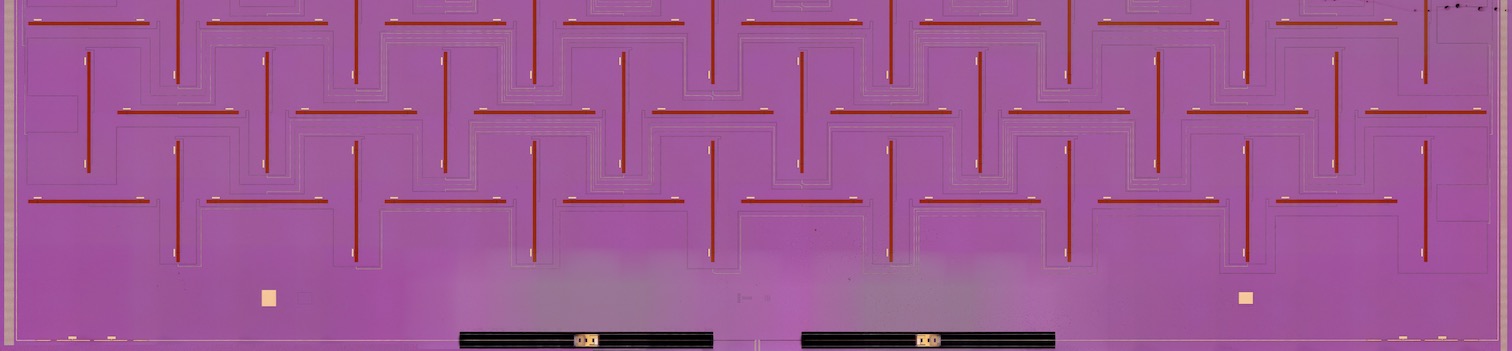}
\caption{Photo of the edge of a 40\,GHz pixel, with bolometers and filters visible. The two orientations of slot antennas separately receive the two linear polarizations.  The power from each set of slots is collected by the microstrip summing trees, passes through the band-pass filter on the lower corners of the photo, and thermalizes in the two TESs on the bottom.  The two filters and bolometers separately receive the two linear polarizations.(color version online)}\label{Antenna2Bolometerphoto}
\end{center}
\end{figure}

\begin{figure}[htbp]
\begin{center}
\includegraphics[width=0.8\linewidth, keepaspectratio]{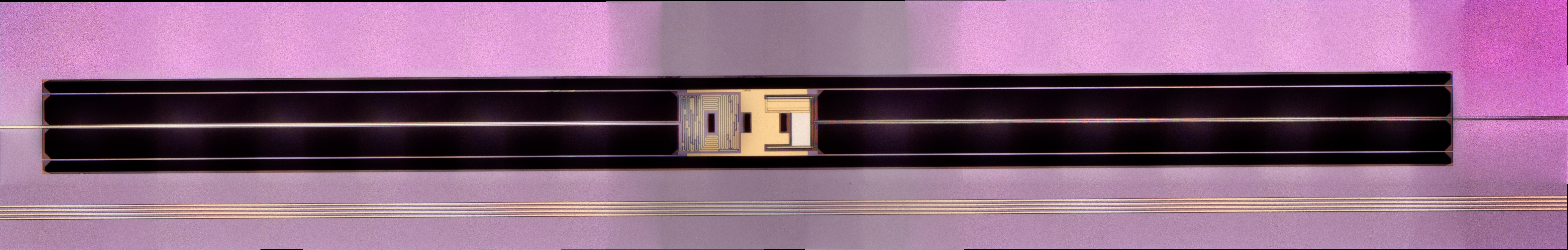}
\caption{Photo of a TES bolometer from a 40\,GHz detector. The bolometer island is suspended from the thermal bath (the detector tile) by six $\sim1.4$\,mm-long legs, forming a weak thermal link. The thickness of the membrane is 1\,$\mu$m. (color version online)}\label{TESphoto}
\end{center}
\end{figure}

As shown in Figure. \ref{Antenna2Bolometerphoto}, the antenna coupled TESs used in BA 30/40\,GHz are following the traditional design of the former BK experiments, operating at bath temperature $T_{bath}$ about $250 ~{\rm mK}$, using two sets of slot antennas oriented in orthogonal directions to measure the polarized linear Stokes parameters. The size of the individual slots is much larger in 30/40\,GHz than in the former higher frequency bands, and the 8$\times$8 slots arrays result in pixel sizes of $\sim$25$\times$25\,mm$^2$ for 30\,GHz and $\sim$20$\times$20\,mm$^2$ for 40\,GHz. In each detector tile we have 4$\times$4 and 5$\times$5 imaging arrays of pixels for 30 and 40\,GHz, respectively. The large pixel size and low detector count impose stringent uniformity and yield requirements across the wafer.

For each polarization, a microstrip summing tree is used to collect and coherently sum the power from all the antenna slots. The final microstrip line passes through a band-pass filter that defines the optical band of the camera, before depositing power on the detector island. Figure. \ref{TESphoto} is a zoom-in photo of one of the 40\,GHz bolometers. The gold meander on the detector island is an open-ended microstrip whose impedance is well matched to the superconducting transmission line. It terminates the end of the microstrip circuit, converts optical power into heat before the reflected EM wave get back out to the transmission microstrip.  The Niobium patterns on the gold meander as shown in the photo are designed to optimize the absorption of the optical power at 30\, and 40\, GHz. There are two TESs connected in series on each detector island- the Ti TES will be used in sky observation while the Al one is used in optical testing thanks to its higher saturation power $P_{sat}$. Detailed information of the detector island structure can be found in [4].

The detector island is suspended by 6 legs, forming the weak thermal link between the island and thermal bath with a thermal conductivity $G$. The length of the legs are chosen in order to meet the photon-noise dominant criteria while avoiding saturation during operation. The fabricated 40\,GHz bolometers (shown in Figure. \ref{TESphoto}) have leg lengths targeting saturation powers 2.5 times of the expected optical loading. We define a safety factor $SF$ as the ratio between saturation power and the optical loading. As a comparison, the leg length of 95\,GHz bolometers used in BICEP3 was roughly half this length. The longer leg gives lower $G$, thereby suppressing the phonon noise to be subdominant to photon noise.  The expected $G$  is $14.5~ {\rm pW/K}$ (at $450~ {\rm mK}$). 

\begin{figure}[htbp]
\begin{center}
\includegraphics[width=0.6\linewidth, keepaspectratio]{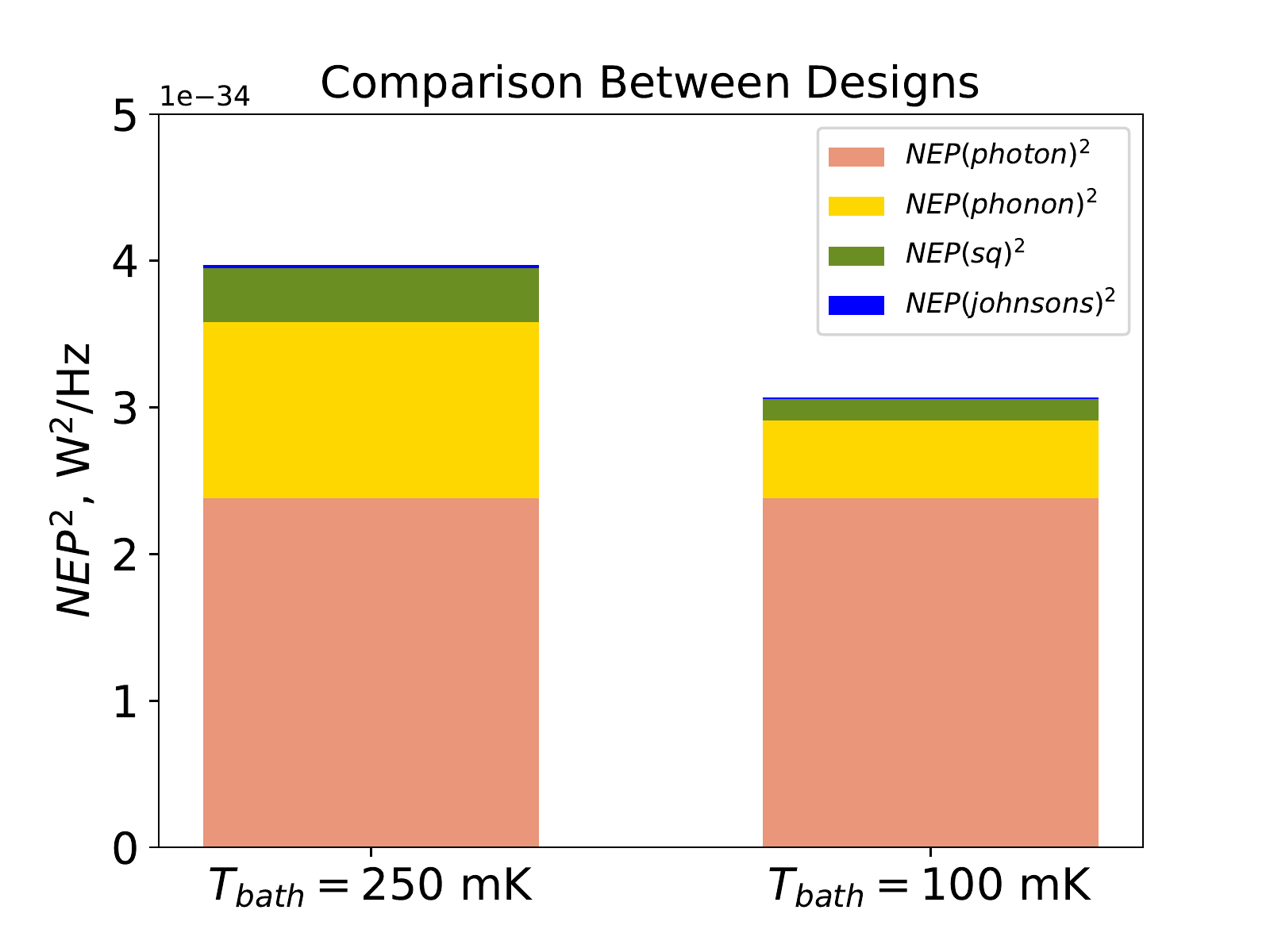}
\caption{NEP contributions of two different 40 GHz bolometer designs, both assuming SF=2.5. {\it Left} Bolometer design with $T_{bath}=250~ {\rm mK}$. It is very close to the bolometer we are currently using for BA, requiring low $G$ bolometer design as described in text. {\it Right} A design with $T_{bath} = 100~ {\rm mK}$. $T_c$ and Operating impedance are also shifted down by 2.5 in this case. While the NEP benefits from a lower bath temperature by a factor of $\sqrt{3.07/3.98} = 0.88$, the expense and effort cannot be justified for this application when weighed against other needs.(color version online)}\label{nep}
\end{center}
\end{figure}

We explored further increasing sensitivity by suppressing the thermal bath temperature $T_{bath}$. The computed noise equivalent power (NEP) contributions from different noise components of 2 different bolometer designs are shown in Figure. \ref{nep}. The left bar is the BA design, with $T_{bath}=250 {\rm mK}$, giving a total NEP  $3.98\times 10^{-17}~{\rm W/\sqrt{Hz}}$. A design operating at $T_{bath} = 100~{\rm mK}$ is present at right. Both cases are assuming $SF=2.5$. The lower bath temperature gives a NEP 12\% lower, but the cost and complication of a more powerful dilution refrigerators required to achieve these lower temperatures is difficult for our team to justify for such marginal gains.

\section{Measured Detector Parameters}

\begin{figure}[htbp]
\begin{center}
\includegraphics[width=0.5\linewidth, keepaspectratio]{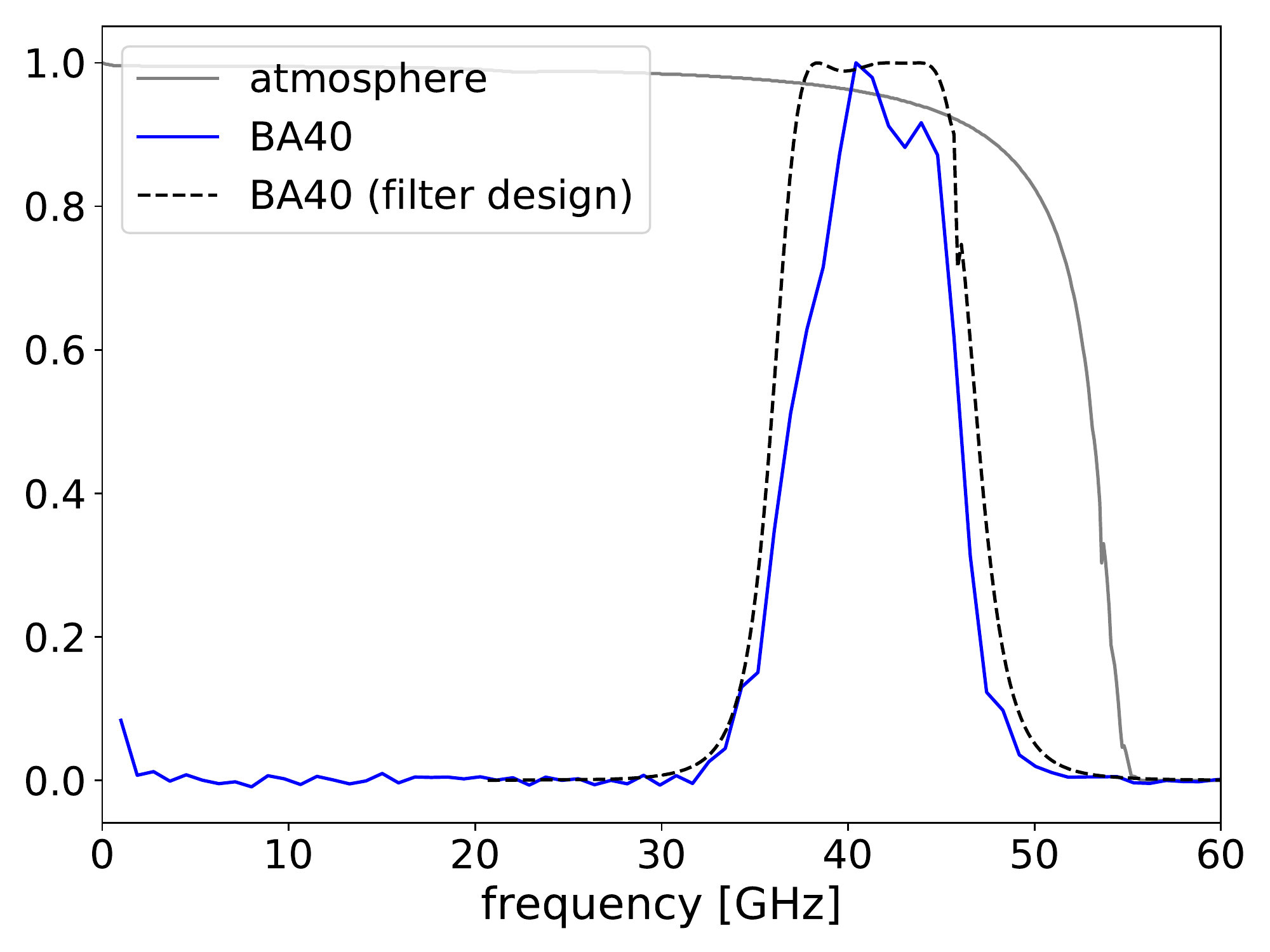}
\caption{Normalized detector spectra and South Pole atmospheric transmission. The {\it blue solid line} shows the averaged 40\,GHz spectrum, with band center at 41.5\,GHz and 28\% band width. The {\it black dash line} shows the designed 40 GHz bandpass filter.  The {\it black solid line} shows typical atmospheric transmission at the South Pole.(color version online)}\label{specsvsatmos}
\end{center}
\end{figure}

We have fabricated and tested several 40\,GHz detector tiles with the same antenna and filter design but different bolometer leg lengths, of which three have the desired bolometer properties (the design is shown in Figure \ref{TESphoto}). Spectra have been measured for the fabricated devices using a Fourier Transform Spectrometer (FTS). The average 40\,GHz spectrum is shown in Figure.\ref{specsvsatmos}, with band center at 41.5\,GHz and $\sim$28\% bandwidth.  For reference, we co-plot the typical atmospheric transmission at South Pole.  

The optical response of the detectors has been measured using two aperture-filling black body sources at room (300\,K) and liquid nitrogen (77\,K) temperatures. These results together with the measured spectra provide an estimate of the end-to-end optical efficiency- roughly 31\%. The expected total optical loading is calculated within our 30/40 GHz band pass. We got 1.0\,pW for 40\,GHz, and 0.5\,pW for 30\,GHz, including loading from CMB and atmosphere, as well as reflective and absorptive losses in the test-bed filter stack and window. The in-band loading from atmosphere has an averaged temperature of 220\,K with absorptivity of 0.06.

\begin{figure}[htbp]
\begin{center}
\includegraphics[width=0.45\linewidth, keepaspectratio]{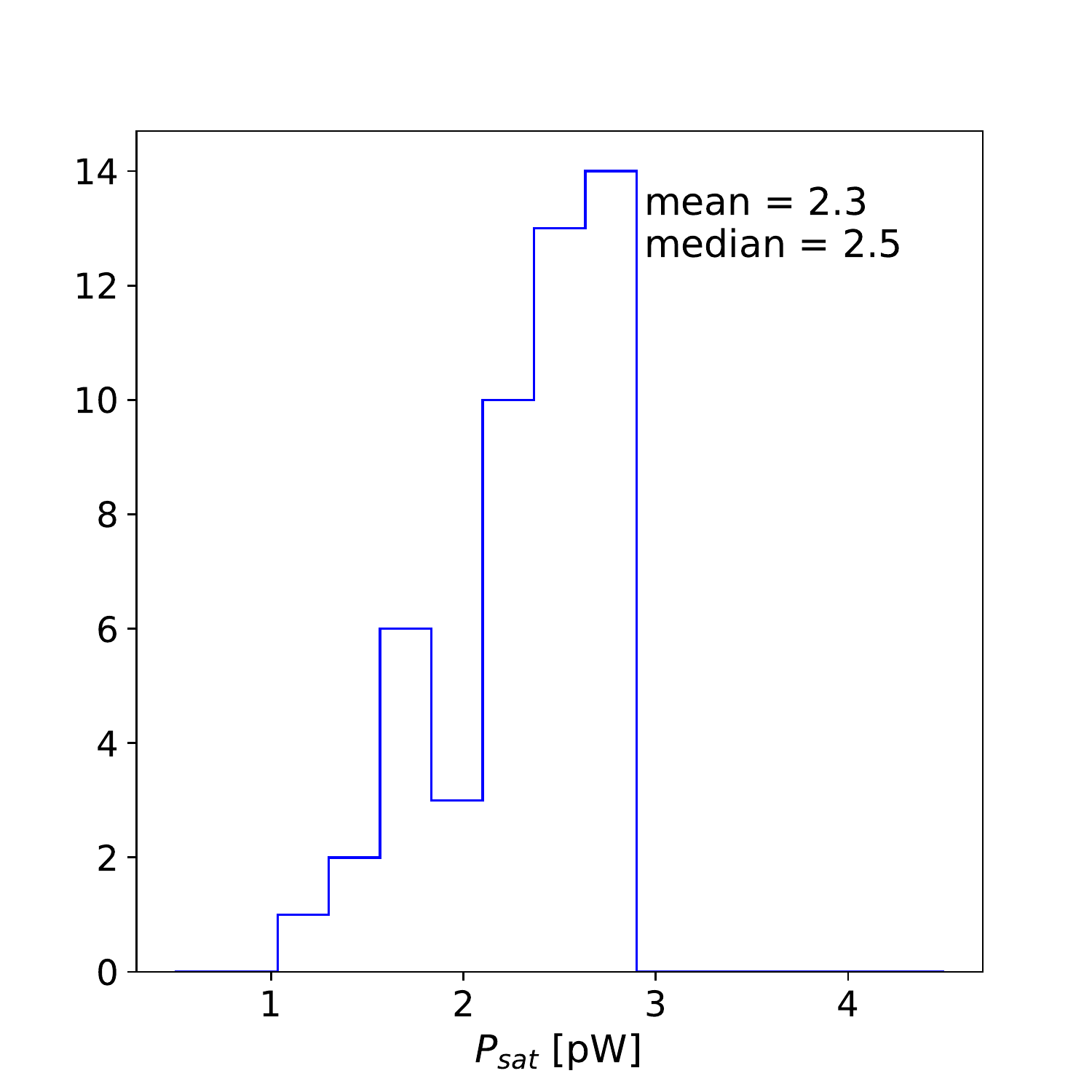}
\includegraphics[width=0.45\linewidth, keepaspectratio]{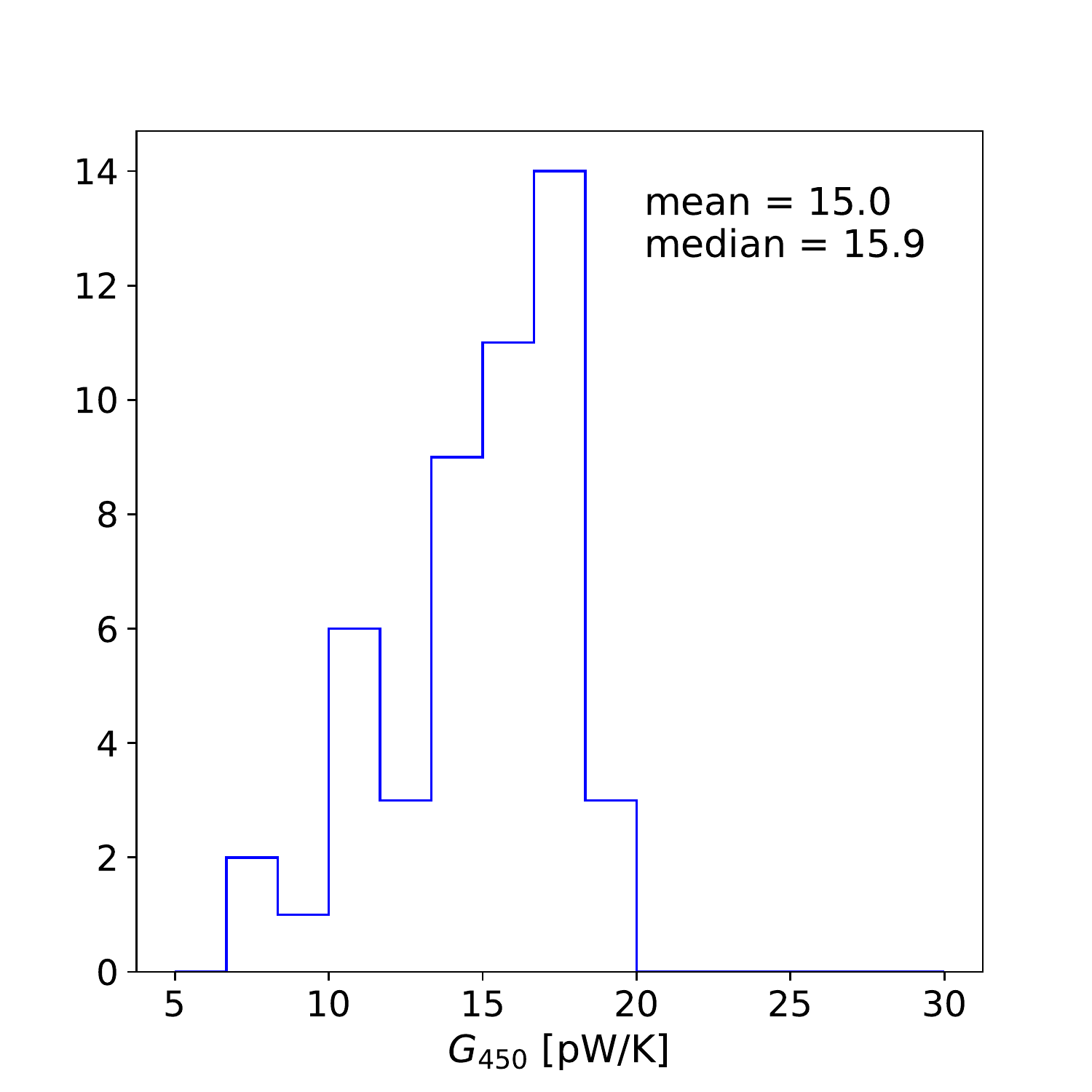}
\caption{Histograms of $P_{sat}$ and $G_{450 }$ of one 40\,GHz tiles. {\it Left,} $P_{sat}$, measured with an operation bath temperature of 275\,mK, has a median of 2.5\,pW with most of the detectors between 1.5 to 3\,pW. {\it Right,} the bolometer island thermal conductivity at 450\,mK $G_{450}$ has a median of 15.9\,pW/K. (color version online)}\label{G450andPsat}
\end{center}
\end{figure}

\begin{figure}[htbp]
\begin{center}
\includegraphics[width=0.48\linewidth, keepaspectratio]{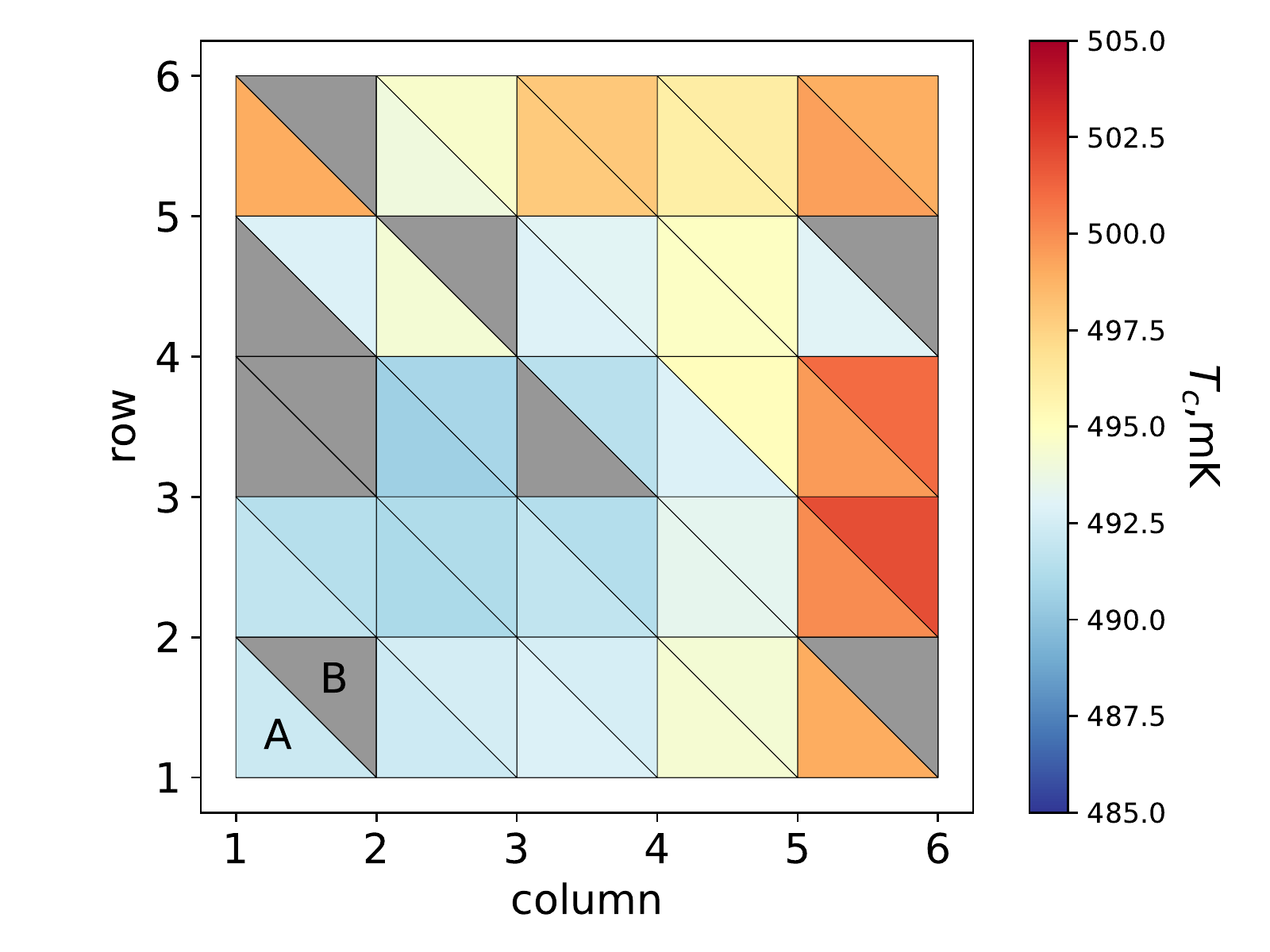}
\includegraphics[width=0.48\linewidth, keepaspectratio]{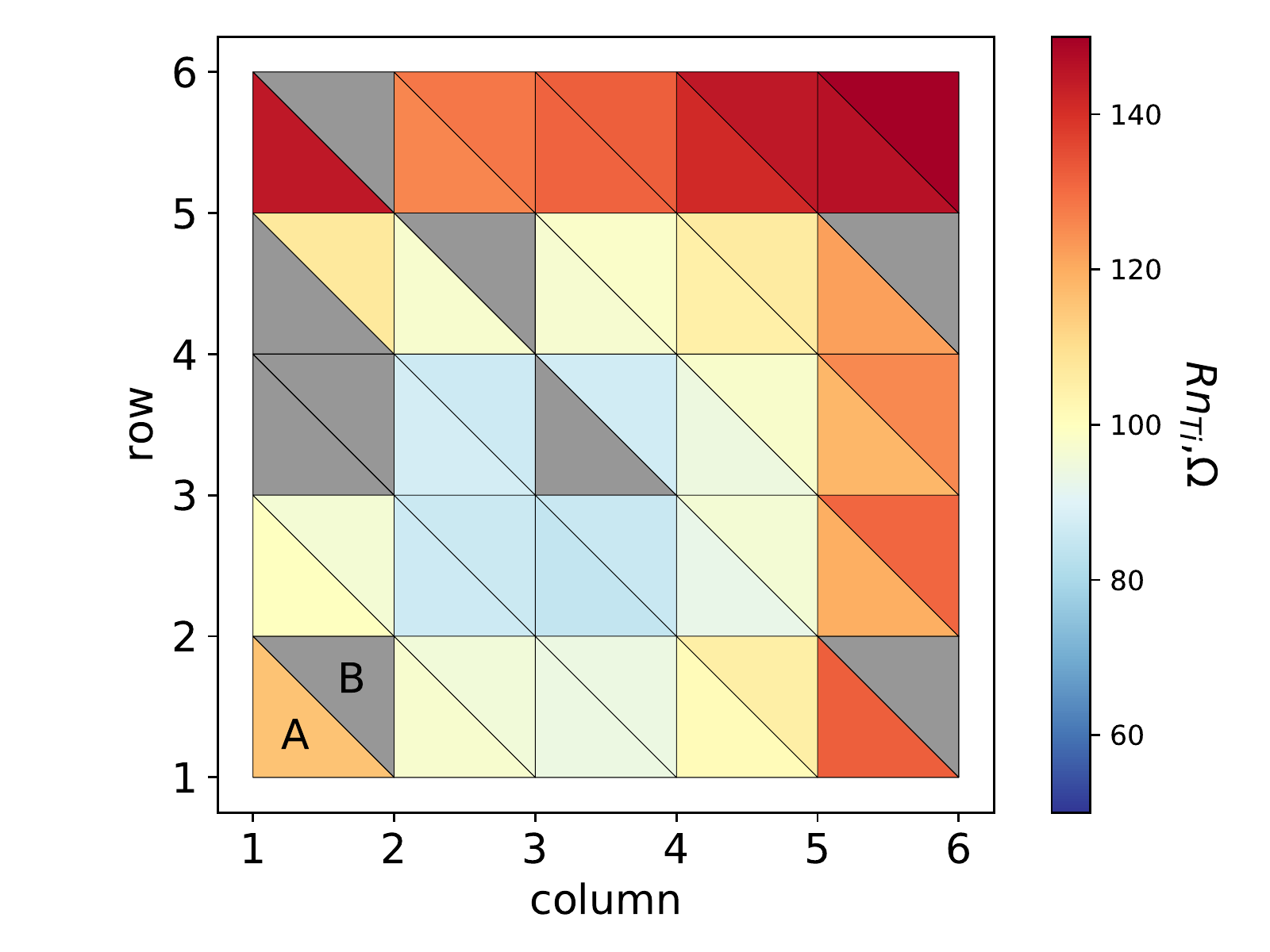}
\caption{The tile map of $T_c$ and $Rn_{Ti}$. The triangles in each square represent the two bolometers for the orthogonal polarization within each pixel. Grey halves represent inoperational  bolometers. $T_c$ is uniform and well controlled to be $\sim$500\,mK. (color version online)}\label{TcandRnti}
\end{center}
\end{figure}

 The left panel of  Figure.\ref{G450andPsat} shows the measured saturation for one 40\,GHz tile operated at $\sim$275\,mK, showing a median value of 2.5\,pW with most of the detectors scattered between 1.5 and 3\,pW. This result matches our design target of $P_{sat}\sim$2.5 times the total optical loading. We have also measured the $P_{sat}$ at a series of $T_{bath}$ to fit the result using $P_{sat} = G_c(T_c^{\beta+1}-T_{bath}^{\beta+1})/T_c^\beta(\beta +1)$. $G_c$ is the thermal conductivity between detector island and thermal bath measured at the TES transition temperature $T_c$, which is $\sim$500\,mK for our Ti TESs. The temperature index $\beta$ is typically 2 for our bolometer design. A histogram of $G$ scaled to 450\,mK with $G_{450} = G_c(450/T_c)^\beta$ is shown in the right panel of Figure. \ref{G450andPsat}, with a median value of 15.9\,pW/K.

Figure. \ref{TcandRnti} shows a wafer map of $T_c$ and the normal resistance of the titanium TESs on one 40\,GHz tile, illustrating spatial achieved uniformity. Such studies are important because we recently switched from fabricating on 100\,mm diameter tiles to 150\,mm diameter tiles, and uniformity is not guaranteed over these larger surfaces.  Non-uniformity in $R_n$ exists as shown in left panel of Figure. \ref{TcandRnti}, which may add difficulty to find a common bias in columns of our SQUID-based time division multiplexing readout.  The thickness of Ti film has been measured over the 150\,mm wafer. The variance between edge and center is about 15\% of the average value, which is not enough to explain the large scattering of $R_n$. We think the variance could be caused by some contact problems that exist within or between layers. Efforts have been made in modifying fabrication process to fix this problem in the future. For the upcoming season, we took noise measurements at multiple bias points and confirmed the fabricated devices can work stably with photon noise dominant in a range of common bias points with operating resistance between 40 and 75\, m$\Omega$.


\section{Background Noise Measurements and Expected Sensitivity}

\begin{figure}[htbp]
\begin{center}
\includegraphics[width=0.8\linewidth, keepaspectratio]{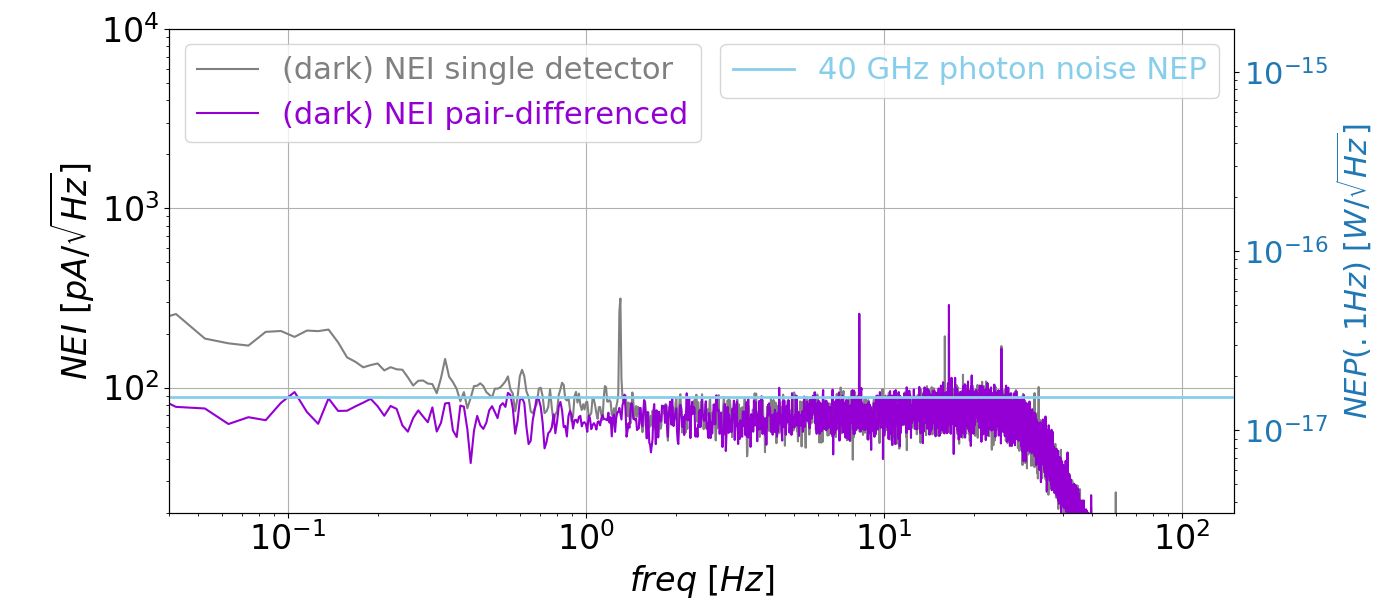}
\caption{The current noise spectrum of a 40\,GHz dark detector measured in a covered configuration. The grey line is the spectrum of an individual bolometer and the purple line is the difference spectrum of between 2 detectors of polarization A and B in same pixel, divided by $\sqrt{2}$ to aid comparison to the single detector spectrum.  The suppressed low frequency noise demonstrates that the 1/f effects are common mode and thus removable.  The light blue line shows the expected photon noise level of $1.54\times10^{-17}~{\rm W/\sqrt{Hz}}$. The estimated total NEP including the measured background noise and photon noise is $2.07\times10^{-17}~{\rm W/\sqrt{Hz}}$. (color version online)}\label{noisespec}
\end{center}
\end{figure}

We fabricate some detectors on our tiles without antennas to function as dark bolometers and provide a baseline for noise performance.  Figure. \ref{noisespec} shows the noise spectrum of one such detector, but with the tile module covered to further limit optical stimulation. The grey and purple lines represent the dark noise spectra of an individual detector and of the pair-difference time stream between two detectors in the same pixel. The estimated photon noise level according to expected optical loading at 40\,GHz is given in light blue. Adding the measured noise at 0.1 Hz with estimated photon noise of $1.54\times10^{-17}~{\rm W/\sqrt{Hz}}$, we expect a total NEP $2.07\times10^{-17}~{\rm W/\sqrt{Hz}}$. Given that the science band is between 0.05 and 1\,Hz in BK with a telescope scan rate of 2.8\,deg/s, this plot is a preliminary validation of our photon-noise dominated design.

\section{Outlook}
We have demonstrated our design goal of photon-noise dominated 40\,GHz devices for use in the first BA receiver. Now we are half-way through the fabrication of 40\,GHz tiles, and the rest are expected to come in the following month. The design of 30\,GHz detector tiles has been finished, with similar considerations from the 40\,GHz development process. Experience and knowledge accumulated in 40\,GHz gives us confidence in a successful 30\,GHz development as well. The first 30\,GHz tile is currently under fabrication and is expected to be tested by Fall 2019.  We expect that after 3 full years of observations, all four BICEP Array cameras will measure primordial gravitational waves to a precision σ(r) between 0.002 and 0.004, depending on foreground complexity and the degree of lensing removal.

\begin{acknowledgements}
The BICEP/Keck project have been made possible through a series of grants from the National 
Science Foundation including 0742818, 0742592, 1044978, 1110087, 1145172, 1145143, 1145248, 1639040, 1638957, 1638978, 1638970, \& 1726917 and by the Keck Foundation.
The development of antenna-coupled detector technology was supported by the JPL Research and Technology Development Fund and NASA Grants 06-ARPA206-0040, 10-SAT10-0017, 12-SAT12-0031, 14-SAT14-0009 \& 16-SAT16-0002.
The development and testing of focal planes were supported by the Gordon and Betty Moore Foundation at Caltech.
Readout electronics were supported by a Canada Foundation for Innovation grant to UBC.
The computations in this paper were run on the Odyssey cluster supported by the FAS Science Division Research Computing Group at Harvard University.
The analysis effort at Stanford and SLAC is partially supported by the U.S. DoE Office of Science.
We thank the staff of the U.S. Antarctic Program and in particular the South Pole Station without whose help this research would not have been possible.
Tireless administrative support was provided by Kathy Deniston, Sheri Stoll, Irene Coyle, Donna Hernandez, and Dana Volponi. 
\end{acknowledgements}


\end{document}